\documentclass[amsfonts,amssymb,aps,amsmath,twoside,showpacs]{revtex4}
\usepackage{graphicx}
\usepackage{dcolumn}
\usepackage{bm}
\usepackage{times}

\begin{document}
\bibliographystyle{apsrev}

\title{Quantum-classical transitions in Lifshits tails with magnetic fields}

\author{Hajo Leschke} 
\affiliation{%
Institut f\"ur Theoretische Physik, Universit\"at Erlangen-N\"urnberg, 
Staudtstr.\ 7, 91058 Erlangen, Germany
}%
\author{Simone Warzel}%
\affiliation{%
Institut f\"ur Theoretische Physik, Universit\"at Erlangen-N\"urnberg,
Staudtstr.\ 7, 91058 Erlangen, Germany
}%


\begin{abstract}
  We consider  Lifshits' model of a quantum particle subject to a repulsive Poissonian random potential and
  address various issues related to the influence of 
  a constant magnetic field on the 
  leading low-energy tail of the integrated density of states. 
  In particular, we propose the magnetic analog of a $ 40 $-year-old landmark result of Lifshits for short-ranged single-impurity
  potentials $ U $. The Lifshits tail is shown to change its character 
  from purely quantum, through quantum-classical, to purely classical with increasing range of $ U $. 
  This systematics is explained by the increasing importance of the 
  classical fluctuations of the particle's potential energy in 
  comparison to the quantum fluctuations associated with its kinetic energy. 

\end{abstract}

\pacs{71.23.An, 73.43.Nq \hfill To appear in slightly different form in \emph{Physical Review Letters}}
\maketitle

\section{Introduction}
The exponential distortion of van Hove singularities of the (integrated) density of states (IDOS) near band edges is 
a fundamental feature of disordered electronic systems.
The associated leading (band-edge) falloff of the IDOS is commonly referred to as a Lifshits tail.
For an unadulterated theoretical understanding of this phenomenon, Lifshits studied
an idealized statistical model of a quantum particle 
in three-dimensional configuration space $ \mathbb{R}^3 $ subject  
to macroscopically many repulsive impurities which are distributed completely at random \cite{Lif63}. 
Within this model the (low-energy) falloff of the IDOS originates in exponentially rare realizations of the randomness with 
large impurity-free regions, 
where the particle's potential energy is solely
due to the impurities outside. It therefore depends on the range of the impurities.   
Lifshits argued that all impurities of short range create the same tail (universally given by  (\ref{LifshitsB=0}) 
below with $ d = 3 $). 
A mathematical proof of this result turned out to be difficult \cite{DonVar75,Pas77,Nak77}. 
It was achieved with the help of Donsker and Varadhan's celebrated large-deviation theorem 
for the long-time asymptotics of certain Wiener path integrals  \cite{DonVar75}. 
Shortly after, Pastur observed that 
the Lifshits tail ceases to be universal in case of long-ranged impurities, but 
rather depends on details of the potential created by a single impurity \cite{Pas77}.

Apart from its obvious relevance to highly doped semiconductors, the phenomenon of Lifshits tailing is of interest  
for a variety of other disordered systems. 
An example is Brownian motion in random media for which the long-time survival probability 
is related to the low-energy behavior of the
IDOS by Laplace transformation and a Tauberian theorem \cite{DonVar75,KaHu84,Szn98}. 
Another example is the random-bond Ising model exhibiting Griffiths singularities \cite{Nie89}.    
The basic large-deviation mechanism responsible for the creation of Lifshits 
tails is also claimed to be the reason for the suppression of 
superconductivity in systems with magnetic impurities \cite{BaTr97} 
and for the disorder-induced rounding of certain quantum phase transitions \cite{Voi03}.

In the present paper we report on a number of new theoretical, mostly rigorous results on the fate of Lifshits tails 
in a constant magnetic field. 
Rigorous studies of Lifshits' model for a two-dimensional configuration space $ \mathbb{R}^2 $ have already revealed that 
the presence of a magnetic field brings about remarkable changes 
in comparison to the nonmagnetic case \cite{BHKL95,Erd98,HuLeWa99,Erd01}. 
In $ \mathbb{R}^3 $ an additional feature comes into play: apart from universal and nonuniversal Lifshits 
tails of purely quantum and purely  classical character, respectively, there exists a wide class of tails with  
coexistence of both characters. They occur for impurities of intermediate range.  
Our main goal is to develop the physical heuristics behind these results for $ \mathbb{R}^2 $ and $ \mathbb{R}^3 $. 
Hereby the new facet lies in both, the inclusion of 
a magnetic field and the consideration of non-short-ranged impurities. 
Mathematical proofs for the case $ \mathbb{R}^3 $ will be published elsewhere.

\section{Model} 
Lifshits' model concerns a spinless particle with mass $ m > 0 $
and electric charge $ q \neq 0 $, which we will suppose to move in $ d $-dimensional Euclidean space $ \mathbb{R}^d $. 
Its total energy is represented by a
random Schr\"odinger operator on the Hilbert space $ L^2(\mathbb{R}^d) $ which is informally defined as
\begin{equation}\label{eq:HamOp}
  H(V) := \frac{1}{2m} \sum_{k=1}^d \big( -i \hbar \frac{\partial\;}{\partial x_k} - q A_k \big)^2  + V. 
\end{equation}
Here  $ 2 \pi \hbar > 0 $ denotes Planck's constant, $ - i \hbar \partial/\partial x_k $ the $ k $th component of the 
canonical-momentum operator, and $ A_k $ the $ k $th component of  
a vector potential $ A : \mathbb{R}^d \to \mathbb{R}^d $ 
describing a constant magnetic field of strength $ B \geq 0 $.
Repulsive impurities generate
the Poissonian random potential $ V : \mathbb{R}^d \to \mathbb{R} $ informally given by 
\begin{equation}\label{eq:poisson}
  V(x) := \sum_j\, U\big(x-p_j\big), \quad U \geq 0.
\end{equation}
For a fixed realization
of the randomness,
the point $p_j \in \mathbb{R}^d $ stands for the position of the
$j$th impurity repelling the particle at $x \in \mathbb{R}^d $ through a nonrandom, nonnegative
single-impurity potential $ U: \mathbb{R}^d \to \mathbb{R} $, which we assume to be integrable, square-integrable,  
and strictly positive 
on some nonempty open subset of $ \mathbb{R}^d $ \cite{note1}.  
The impurity positions are independently, identically, and uniformly 
distributed throughout $ \mathbb{R}^d $ with mean concentration $ \varrho > 0 $ such that 
the probability of finding $J \in \{0,1,2,\dots\}$ 
impurities in a region $ \Lambda \subset \mathbb{R}^d $ of volume $ | \Lambda | := \int_{\Lambda} d^d x $
is given by Poisson's law 
$  \exp[- \varrho | \Lambda | ] \left(\varrho | \Lambda | \right)^{J}/ J! $.
Denoting the corresponding probabilistic (ensemble) average by an overbar, 
the IDOS resulting from (\ref{eq:HamOp}) with (\ref{eq:poisson})
at a fixed energy $ E \in \mathbb{R} $
can be defined \cite{LeMuWa03} as
\begin{equation}\label{DefIDOSinformell}
 N(E) := \overline{\big\langle x | \Theta\big( E - H(V)\big) | x \big\rangle},
\end{equation} 
in terms of Heaviside's (left-continuous) 
unit-step function  $ \Theta $. 
Thanks to unitary invariance of the kinetic-energy operator $ H(0) $ under magnetic translations and due to  
the $ \mathbb{R}^d $-homogeneity of the Poissonian potential, $ N(E) $ is independent of the chosen $ x \in \mathbb{R}^d$ labeling the
position representation.
By the decay of $ U $ at infinity the half-line $ [ \varepsilon_0 , \infty [ \subset \mathbb{R} $ is not only the set of growth points of the function 
$ N : \mathbb{R} \to \mathbb{R} $ but also coincides with the spectrum of $ H(V) $ almost surely, that is, with probability one \cite{note2}.
Here $  \varepsilon_0 \geq 0 $ denotes the ground-state 
energy of $ H(0) $, which is zero for $ d = 1 $ and equal to the lowest Landau-level energy $ \hbar | q | B / 2 m $ 
for $ d= 2 $ and $ d= 3 $.

\section{Quantum-classical transitions} 
At energies $ E \downarrow \varepsilon_0 $, the particle will be localized \cite{Sto95}
in a large region $  \Lambda_0 \subset \mathbb{R}^d $ without impurities.
If $ U $ is short-ranged, its potential energy in $  \Lambda_0 $ is to a good approximation zero.  
By the spatial confinement its kinetic energy is not smaller than the lowest eigenvalue of $ H(0) $ 
when the latter is Dirichlet restricted to $ \Lambda_0 $.  
Lifshits suggested that at low energies $ N(E) $ 
will be determined by the region $  \Lambda_0(E) \subset \mathbb{R}^d $
with the smallest volume $ |\Lambda_0(E)| $ for which the lowest Dirichlet eigenvalue of $ H(0) $  
coincides with the given $ E $ \cite{FrLu75}. 
He therefore proposed the following asymptotic formula \cite{note3} for the leading low-energy falloff 
of the IDOS as $ E \downarrow \varepsilon_0 $
\begin{equation}\label{eq:eq:approxN}
       \log N(E)  \sim  \log \, \text{Prob}\big\{ \text{$\Lambda_0(E) $ is free of impurities}\big\} 
        = - \varrho \, \left| \Lambda_0(E) \right|
\end{equation} 
if $ U $ is short-ranged. If $ U $ is long-ranged, the particle inside $ \Lambda_0 $ 
acquires a potential energy due to the long-distance decay of potentials $ U $ generated by impurities
located outside $ \Lambda_0 $, that is, in $\mathbb{R}^d  \backslash \Lambda_0 $. 
Given the impurity-free region $ \Lambda_0 $, this potential energy is on average of the order of magnitude
\begin{equation}\label{eq:potenergy}
  \varrho \, \int_{\mathbb{R}^d \backslash \Lambda_0 } \mkern-20mu d^d x \; U(x). 
\end{equation}
Supposing that  $ U $ varies slowly on the scale of the particle's de Broglie wavelength, the kinetic energy of the particle inside $ \Lambda_0 $ will 
still be given approximately by the lowest Dirichlet eigenvalue of $ H(0) $. 
Therefore a basic question is, whether this kinetic energy, caused by the spatial confinement to $ \Lambda_0 $, dominates 
(\ref{eq:potenergy}) or not as $ | \Lambda_0 | \to \infty $.
If yes, the Lifshits tail has a purely quantum character and is universally given by (\ref{eq:eq:approxN}). 
If not, it will in general depend on details of the decay of $ U $ and exhibit classical features. 
Moreover, if the quantum fluctuations related to the kinetic energy can be neglected completely, the Lifshits tail has  
a purely classical character in the sense that 
\begin{equation}\label{eq:classical}
  \log N( \varepsilon_0 + E) \sim \log N_{cl}(E)
\end{equation}
as $ E \downarrow 0 $. Here 
\begin{equation}
N_{cl}(E) := \left(\frac{ m}{2 \pi \hbar^2}\right)^{d/2} \frac{ \overline{\big( E - V(0)\big)^{d/2} \Theta\big( E - V(0) \big)} }{ \Gamma(1+d/2)}          
\end{equation}
is the (quasi-) classical IDOS \cite{Kan63,LeMuWa03} with $ \Gamma $ denoting 
Euler's gamma function. 
In accordance with a theorem of Bohr and van Leeuwen on the nonexistence of diamagnetism in classical physics,   
$ N_{cl} $ is independent of the magnetic field.

\section{Case $ \mathbf{ B = 0 }$}
It will be instructive to briefly recall what happens in the zero-field case. 
Here the isoperimetric inequality of Strutt (= Rayleigh), Faber, and Krahn \cite{note4}
shows that
balls have the smallest volume for a given lowest Dirichlet eigenvalue of $ H(0) $.
Moreover, the volume $  \left| \Lambda_0  \right| $ of a ball $  \Lambda_0 $ whose associated lowest Dirichlet 
eigenvalue is $ E_0(\Lambda_0) $ can be inferred from a scaling argument:
\begin{equation}\label{eq:eigenwertB=0}
  E_0(\Lambda_0) = \frac{\kappa_d \hbar^2}{2m}\,  \left| \Lambda_0 \right|^{-2/d}.
\end{equation}
Here $ \kappa_d $ is the lowest eigenvalue of the negative Laplacian when Dirichlet restricted to
a ball in $ \mathbb{R}^d $ of unit volume, for example, $ \kappa_1 = \pi^2 $, $\kappa_2 = \pi \xi_0^2 $, 
with $ \xi_0 = 2{.}404\dots $ being the smallest 
positive zero of the zeroth Bessel function of the first kind, 
and $ \kappa_3 = \pi^2 \left( 4 \pi /3 \right)^{2/3} $.
Combining (\ref{eq:eq:approxN}) and (\ref{eq:eigenwertB=0}) one obtains Lifshits' landmark result \cite{Lif63} 
for the leading low-energy falloff of the IDOS 
as $ E \downarrow 0 \, (= \varepsilon_0 ) $ if $ U $ is short-ranged:
\begin{equation}\label{LifshitsB=0}
  \log N(E) \sim  - \varrho \left(\frac{\kappa_d \hbar^2 }{2 m E}\right)^{d/2}. 
 \end{equation}
As an aside, we note that (\ref{LifshitsB=0}) with $ d = 1 $ remains valid \cite{Egg72} 
in the limiting case of point impurities \cite{note1}. 
If $ U $ is long-ranged in the sense that it 
has an (integrable) algebraic decay proportional to $ | x |^{-\alpha} $ as $ | x | \to \infty $ 
with some exponent $ \alpha\, (> d) $, the 
potential energy (\ref{eq:potenergy}) is proportional to $| \Lambda_0 |^{1- \alpha/d} $.
As $ | \Lambda _0 | \to \infty $, it is therefore negligible in comparison to the kinetic energy (\ref{eq:eigenwertB=0})
if and only if $ \alpha > d + 2 $. 
More generally, if the decay is faster than algebraic with exponent $ d + 2 $, 
the Lifshits tail was proven \cite{DonVar75,Pas77,Nak77} to be universally given by (\ref{LifshitsB=0}). 
If $ \alpha < d + 2 $ the total energy is dominated by the potential energy and the Lifshits tail
has indeed a purely classical character in the sense that (\ref{eq:classical}) holds \cite{Pas77}.
Algebraic decay with exponent $ \alpha = d + 2 $ therefore discriminates between Lifshits tails of purely quantum and 
those of purely classical character if $ B = 0 $. 
In this borderline case, $ \alpha = d + 2 $, coexistence of both quantum and classical behavior is expected \cite{LeMuWa03}.

\section{Case $ \mathbf{B > 0} $}
What changes when a constant magnetic field is turned on?
First of all, a magnetic field of strength $ B $ introduces 
the length scale $ \ell := \sqrt{ \hbar / | q | B } $ and the energy scale
$ \hbar^2/ 2m \ell^2 $ ($= \varepsilon_0 $ for $ d = 2$ and $ d= 3 $). 
Of course, (\ref{eq:eq:approxN}) continues to hold in the short-ranged case.
It is the shape and mainly the volume of the region $ \Lambda_0(E) $ through which the magnetic field enters.
Physical intuition suggests that an external magnetic field favors localization effects. 
This implies that
the energy of a particle which is confined to some region is dramatically diminished in comparison 
to the case $ B = 0 $.
To discuss this in more detail, it is helpful to consider first 
the (idealized) Quantum Hall situation with the 
particle and all impurities confined to a plane $ \mathbb{R}^2 $ perpendicular to the constant magnetic field.

\subsection{Case $ \mathbf{B > 0 }$ and $ \mathbf{d = 2} $}
Due to the rotational symmetry about the magnetic-field direction 
it is plausible 
that balls in $ \mathbb{R}^2 $, that is disks, still yield the smallest area 
for a given lowest eigenvalue of $ H(0) $.
The underlying magnetic isoperimetric inequality was proven in \cite{Erd96}.
Moreover, the increase of the kinetic ground-state energy $ E_0(\Lambda_0) - E_0(\mathbb{R}^2) = E_0(\Lambda_0) - \varepsilon_0 $
by spatial confinement 
to a large disk $ \Lambda_0 \subset \mathbb{R}^2 $  
with area $ | \Lambda_0 | $ is
asymptotically given by \cite{Erd98}
\begin{equation}\label{eq:kinenerg2d}
 E_0(\Lambda_0) - \varepsilon_0 = \varepsilon_0 \, 
  \exp\!\left[ - \frac{| \Lambda_0 |}{2 \pi \ell^2}\big( 1 + o(1) \big)\right],
\end{equation}
where ``little oh'' $ o(1) $ tends 
to zero as $ | \Lambda_0 | \to \infty $. 
The exponential dependence on the area $ | \Lambda _0 | $ 
is a consequence of the fact that the circularly symmetric ground-state wave function of the 
infinite-area kinetic-energy operator $ H(0) $ for $ B > 0 $ is (in contrast to the case $ B = 0 $) square-integrable 
and even exponentially localized. 
For short-ranged $ U $ a combination of (\ref{eq:eq:approxN}) and (\ref{eq:kinenerg2d}) yields 
a power-law falloff of the IDOS
near the (almost sure) ground-state energy $ \varepsilon_0 > 0 $ of $ H(V) $ in the sense that
\begin{equation}\label{eq:2dheur}
  \log  N(\varepsilon_0 + E )  \sim  
  \log  E^{ 2 \pi \varrho  \ell^2}  \sim -  2 \pi \varrho  \ell^2 \, | \log  E | 
\end{equation} 
as $E \downarrow 0$. 
This stands in sharp contrast 
to the exponential falloff (\ref{LifshitsB=0}) if $ B = 0 $.
Given (\ref{eq:eq:approxN}), the difference is due to the fact that the finite-area kinetic ground-state energy 
(see (\ref{eq:kinenerg2d}) and (\ref{eq:eigenwertB=0}), respectively) approaches its infinite-area limit $ \varepsilon_0 $
exponentially if $ B > 0 $ but only algebraically if $ B = 0 $, as the 
disk $  \Lambda _0 $ is blown up to exhaust all of the plane $ \mathbb{R}^2 $.  
Depending on whether the exponent $ 2 \pi \varrho \ell^2 $ in (\ref{eq:2dheur}), which is just the mean number of impurities 
in a disk of radius $ \sqrt{2} \ell $, is smaller or larger than one, the IDOS exhibits a root-like or true power-law falloff. 
The resultant divergence of the 
DOS $ d N / dE $ at $ \varepsilon_0 $ if $ 2 \pi \varrho \ell^2 < 1 $ should be observable in suitable experiments. 
We note that in the limiting case of point impurities \cite{note1} the lowest-Landau-band approximation to $ N $ is known 
exactly \cite{BeGrIt84} with a Lifshits tail (see also \cite{Fur00}) differing from (\ref{eq:2dheur}).

A nontrivial mathematical proof of (\ref{eq:2dheur}) was given by Erd\H{o}s \cite{Erd98} 
for $ U $ with compact support.  
Building on his result, (\ref{eq:2dheur})  was shown to hold for any $ U $
which decays faster than any Gaussian at infinity \cite{HuLeWa99}.
In fact, this is plausible from the heuristic point of view. 
When estimating the potential energy of a particle in a large impurity-free disk $ \Lambda_0 \subset \mathbb{R}^2 $ by 
(\ref{eq:potenergy}), it turns out to be negligible in comparison to the 
increase of the kinetic energy  given by (\ref{eq:kinenerg2d}) if and only if $ U $ decays faster than any Gaussian.
Conversely, if $ U $ decays slower than any Gaussian, the Lifshits tail is dominated by the potential energy and 
hence of classical character in the sense that (\ref{eq:classical}) holds \cite{BHKL95,HuLeWa99}. 
The discriminating decay of $ U $ for the quantum-classical transition 
is therefore Gaussian if $ B > 0 $ and not algebraic (as in the case $ B= 0 $). 
In the borderline case of Gaussian decay quantum and classical behavior coexist \cite{HuLeWa99,Erd01}.

\subsection{Case $\mathbf{ B > 0} $ and $\mathbf{ d = 3} $}
In contrast to the two-dimensional situation, the presence of a constant magnetic field 
in $ \mathbb{R}^3 $ introduces an anisotropy.
Here the isoperimetric problem of finding those regions which yield the smallest volume 
for a given lowest Dirichlet eigenvalue of $ H(0) $ seems to be unsolved. 
It is natural to assume that its solution is found among convex regions which are axially symmetric about the magnetic-field 
direction.
Assuming right circular cylinders as the solution, one may argue as follows.
For a large confining cylinder $ D \times I  \subset\mathbb{R}^3 $ 
with base disk $ D \subset  \mathbb{R}^2 $ and altitude interval $ I \subset \mathbb{R} $ 
parallel to the magnetic-field direction,  
the increase of the kinetic ground-state energy  
$ E_0(D \!\times\! I)  - \varepsilon_0 $ 
is just a sum of two terms in accordance with (\ref{eq:kinenerg2d}) and (\ref{eq:eigenwertB=0}): 
\begin{equation}\label{eq:kinenerg3d0}
    E_0\big(D \!\times\! I\big)  - \varepsilon_0 
   =  \varepsilon_0 \,\, 
  \exp\!\left[ - \frac{|D|}{2 \pi \ell^2}\big( 1 + o(1) \big)\right] 
        + \frac{ \pi^2 \hbar^2}{2 m |I|^2}.
\end{equation}
As a consequence, among all right circular cylinders the one (to be denoted as $ \Lambda_0 \subset \mathbb{R}^3 $) 
which yields the smallest volume for a given lowest Dirichlet eigenvalue of $ H(0) $,
can be inferred asymptotically from the equation
\begin{align}\label{eq:kinenerg3d}
  E_0(\Lambda_0) - \varepsilon_0 \, & = \inf_{|I| > 0} \,\, E_0\big( (\Lambda_0 / I ) \times I \big) - \varepsilon_0  \notag \\
    & =   \frac{\pi^2 \hbar^2 }{2m}  
            \left( \frac{2 \pi \ell^2 }{| \Lambda_0 |}  
              \log \left| \Lambda_0 \right|^2 \right)^2  \big( 1 + o(1) \big). 
\end{align}
Inserting this result into (\ref{eq:eq:approxN}), we conclude that for short-ranged impurities
the IDOS drops down to zero
near the ground-state energy $ \varepsilon_0 > 0 $ 
of $ H(V) $ according to 
\begin{equation}\label{eq:3dheur}
  \log  N( \varepsilon_0 + E ) \sim  
   \left( \log E^{ 2 \pi \varrho^{2/3} \ell^2} \right)  \,  \varrho^{1/3} \,
  \left( \frac{\pi^2 \hbar^2 }{2m E}\right)^{1/2} 
   \sim - 2 \pi \varrho \ell^2 \, \big| \log E \big| \left( \frac{\pi^2 \hbar^2 }{2m E}\right)^{1/2} 
\end{equation}
as $E \downarrow 0$. 
The rhs is the product \cite{note4a} of 
the rhs of (\ref{eq:2dheur}) and (\ref{LifshitsB=0}) 
with $ d = 1 $, provided one notes that $ \varrho $ in (\ref{eq:3dheur}) is the mean bulk concentration. 
The dominant second factor may be attributed to the effective zero-field motion of the particle 
parallel to the magnetic field. 
A leading asymptotic behavior proportional to $ E^{-1/2} \log E $ 
was also suggested  \cite{Her95} in case of point impurities \cite{note1} for the DOS within the lowest-Landau-band approximation. 

So far we do not have a complete mathematical proof of (\ref{eq:3dheur}), the magnetic analog of Lifshits' $ 40 $-year-old  
result (\ref{LifshitsB=0}) (with $ d = 3 $).
We have a lower bound \cite{War01} on the IDOS, which coincides with the so-called optimal-fluctuation formula \cite{note5}  
and has the same leading asymptotics as the rhs of (\ref{eq:3dheur}).
The asymptotics of our upper bound \cite{War01} however dismisses the logarithmic factor. 
To sharpen the upper bound one should extend Erd\H{o}s' proof \cite{Erd98} from $ d = 2 $ to $ d = 3$.

What changes if the impurity potential $ U $ is long-ranged? The potential energy (\ref{eq:potenergy})
of the particle inside $  \Lambda_0 =  D \times I$ is of the same order of magnitude as the sum of two terms 
\begin{equation}\label{eq:potenergy2}
 \varrho \int_{\mathbb{R}^2\backslash D} \mkern-20mu d^2 x_\perp \,  U_\perp(x_\perp) 
  + 
  \varrho \int_{\mathbb{R}\backslash I} \mkern-15mu d x_\parallel \,  U_\parallel(x_\parallel) 
\end{equation} 
containing $ D $ and $ I $ separately. 
Here we have introduced marginal impurity potentials for the directions perpendicular and parallel to the magnetic field, 
$ U_\perp(x_\perp) := \int_\mathbb{R} dx_\parallel \, U( x_\perp, x_\parallel) $ and 
$ U_\parallel(x_\parallel) := \int_{\mathbb{R}^2}\! d^2x_\perp \, U( x_\perp, x_\parallel) $. 
As $ | D | $, $ | I | \to \infty $, each of the two terms of the potential energy in (\ref{eq:potenergy2}) competes with its 
corresponding term of the kinetic energy in (\ref{eq:kinenerg3d0}). 
As a consequence, apart from Lifshits tails with either purely quantum or purely classical character, there emerges 
a wide class of impurity potentials $ U $ 
yielding Lifshits tails with coexistence of these characters.

Of physical relevance
in the context of screening of charged impurities is the example in which $ U $ decays proportional to 
$ \exp\left[ - (|x|/\lambda)^{\beta} (1 + o(1) ) \right] $ as $ |x| = \big(|x_\perp|^2 + x_\parallel^2\big)^{1/2} \to \infty $ 
with some decay length $ \lambda > 0$ and some exponent $ \beta > 0 $. 
Here the potential energy coming from $ U_\parallel $ in  (\ref{eq:potenergy2}) 
is negligible in comparison to the corresponding kinetic energy in (\ref{eq:kinenerg3d0}) as $ | I | \to \infty $.  
However, the analogous assertion concerning the perpendicular directions as $ | D | \to \infty $ is true
if and only if $ \beta > 2 $. In other words, we expect (\ref{eq:3dheur}) to hold as long as
$ U $ decays faster than any Gaussian. If $  \beta < 2 $, the Lifshits tail was proven to be \cite{War01}
\begin{equation}\label{eq:3dgestr}
  \log  N( \varepsilon_0 + E ) \sim - \pi \varrho \lambda^2 \big| \log E \,\big|^{2/\beta} \left( \frac{\pi^2 \hbar^2 }{2m E}\right)^{1/2} 
\end{equation}
as $E \downarrow 0$. 
Like (\ref{eq:3dheur}) it coincides with the product \cite{note4a} of the logarithms of corresponding tails for $ d= 2 $ and $ d=1 $, 
as follows from
(\ref{eq:classical}) (see \cite{HuLeWa99}) and (\ref{LifshitsB=0}), respectively. 
It incorporates (through $ \hbar $ and $ \lambda $) both quantum and classical features.
For the borderline case $ \beta = 2 $ we conjecture in analogy to (\ref{eq:3dgestr}) and \cite{Erd01} that the Lifshits tail is given
by (\ref{eq:3dgestr}) with $ \beta = 2 $ and $ \lambda^2 $ replaced by $ \lambda^2 + 2 \ell^2 $. 
To summarize, in $ \mathbb{R}^3 $ Gaussian decay discriminates between magnetic Lifshits tails with purely quantum and those with coexisting quantum-classical behavior.

A transition from the coexistence regime to the purely classical one can be found, for example, within the class of 
single-impurity potentials 
$ U $ with (integrable) algebraic decay proportional 
to $ | x |^{- \alpha} $ as $ | x | \to \infty $ with some exponent $ \alpha \,(> 3 = d) $. 
Here the particle's potential energy stemming from $ U_\perp $ in (\ref{eq:potenergy2}) 
always dominates the corresponding kinetic energy in (\ref{eq:kinenerg3d0}). 
Since $  U_\parallel $ decays proportional 
to $ |x_\parallel |^{2-\alpha} $ as $ | x_\parallel | \to \infty $, the 
second term in (\ref{eq:potenergy2}) dominates its kinetic counterpart in (\ref{eq:kinenerg3d0}) 
if and only if $ \alpha < 5  $. In the latter case, the Lifshits tail was indeed proven to have a 
purely classical character in the sense that (\ref{eq:classical}) holds \cite{HuKiWa03}.
Algebraic decay with exponent 
$ \alpha = 5  \,(= d + 2) $ therefore discriminates between magnetic Lifshits tails with 
coexisting quantum-classical and those with purely classical character. \\

\section*{Acknowledgment}
We are much indebted to Peter M\"uller (G\"ottingen University) for critical reading of the manuscript.


\begin{thebibliography}{20}
\bibitem{note1}{We exclude point impurities $ U(x) = u_0 \protect \tmspace  +\thinmuskip {.1667em} \delta (x) $, $ u_0 > 0 $. 
  One reason is that for such impurities (\ref{eq:HamOp}) is not unambiguously defined if $ d \geq 2 $. 
  This problem disappears if $ B > 0 $ and $ d= 2 $ within the lowest-Landau-band approximation, see Ref.~\cite{Erd98}.}
\bibitem{note2}{We note that $ N $ is nonnegative, nondecreasing, and left-continuous. 
It is an open mathematical problem whether $N $ is globally continuous or even differentiable for a Poissonian potential. 
In particular, the existence of the DOS $dN/dE $ as a function (not merely as a distribution!) is not assured.}
On the other hand, since $ \lim_{E \downarrow \varepsilon_0} N(E) = 0 $ by the Lifshits tails reported on in the present paper,
$ N $ is continuous at $ \varepsilon_0 $.
\bibitem{note3}{Here and in the following we use the notation $ f(E) \sim g(E) $ as $ E \downarrow E' $ to indicate that 
    $ \lim_{E \downarrow E'} f(E) / g(E) = 1 $. Moreover, we suppress suitable constants ensuring
dimensionless arguments of the natural logarithm ``$\log$'',
because they become 
irrelevant in the considered limit(s).}
\bibitem{note5}{See Eq.\ (1.9) in Ref.\ \cite{LeMuWa03}. 
    In fact, the optimal-fluctuation formula, which 
  summarizes Lifshits' heuristics for more general random potentials very concisely, 
  is often believed to always yield the correct Lifshits tail.}
\end{thebibliography}
\end{document}